\documentclass[pre, showpacs, showkeys, nofootinbib]{revtex4}
\usepackage{latexsym}
\usepackage{amsmath}
\usepackage[dvips]{graphicx}

\begin {document}
\title{Stochastic properties of systems controlled
by autocatalytic reactions II}
\author{L. \surname{P\'al}}
\email[Electronic address:]{lpal@rmki.kfki.hu} \affiliation{KFKI
Atomic Energy Research Institute H-1525 Budapest 114, POB 49
Hungary}

\begin {abstract}
We analyzed the stochastic behavior of systems controlled by
autocatalytic reaction $A + X \rightleftharpoons X + X, \;\; X
\rightarrow B$ provided that the distribution of reacting
particles in the system volume is uniform, i.e. the point model of
reaction kinetics introduced in arXiv:cond-mat/0404402 can be
applied. Assuming  the number of substrate particles $A$ to be
kept constant by a suitable reservoir, we derived the forward
Kolmogorov equation for the probability of finding $n=0,1,\ldots$
autocatalytic particles $X$ in the system at a given time moment.
We have shown that the stochastic model results in an equation for
the expectation value of autocatalytic particles $X$ which differs
strongly from the kinetic rate equation. It has been found that
not only the law of the mass action is violated but also the
bifurcation point is disappeared in the well-known diagram of $X$
particle- vs. $A$ particle-concentration. Therefore, speculations
about the role of autocatalytic reactions in processes of the
"natural selection" can be hardly supported.
\end{abstract}



\maketitle

\section*{Introduction}

As in the previous paper \cite{lpal04} we define the system
$\mathcal S$  as an aggregation of particles capable for
autocatalytic reactions. Symbols $X$ and $A$ are used for
notations of \textit{autocatalytic and substrate particles},
respectively. We denote by $B$ \textit{the particles of
end-product} which do not take part in the reactions. We assume
that the system is "open" for particles $A$, i.e. the number of
particles $A$ is kept constant by a suitable reservoir. However,
the system is strictly closed for the autocatalytic particles $X$.

In this paper we investigate systems which are governed by
reactions
\begin{equation} \label{a1}
A + {X} \stackrel{\mathrm{k_A}} {\longrightarrow} {X} + {X},
\;\;\;\; {X} + {X} \stackrel{\mathrm{k_{X}}} {\longrightarrow} A +
{X}, \;\;\;\; {X} \stackrel{\mathrm{k_{d}}} {\longrightarrow} {B},
\end{equation}
where $\mathrm{k_A}, \;\mathrm{k_X}$ and $\mathrm{k_d}$ are the
rate constants. These reactions became interesting since
convincing speculations were published \cite{parmon02} about the
"natural selection" based on their properties.

The organization of the paper as follows. After brief discussion
of the kinetic rate equation, in Section I we derive and formally
solve the forward Kolmogorov equation for the probability
$p_{n}(t)$ of finding $n=0,1, \ldots \;\; X$ particles in a system
of volume $V$ at time moment $t \geq 0$. Defining the conditions
of the stationarity  we determine the stationary probability
$\lim_{t \rightarrow  \infty} p_{n}(t) = w_{n}$, and analyze its
properties. In Section II we present a modified stochastic model
of the autocatalytic reactions (\ref{a1}) which is capable of
reproducing the solution of the rate equation, however, brings
about such a large fluctuation in the stationary number of $X$
particles, that the mean value loses practically all of its
information content. In Section III we define the lifetime of a
system  controlled by reaction (\ref{a1}), and calculate exactly
the extinction probability, as well as the mean value of the
system lifetime. Finally, in Section IV we summarize the main
conclusions.

\section{Stochastic model}

In order to make comparison, first the well-known results of the
kinetic rate equation are briefly revisited, and then the
stochastic model of the reactions (\ref{a1}) will be thoroughly
analyzed.

\subsection{Rate equation}

Let $m(t)$ be the number of $X$ particles in the volume $V$ of the
system $\mathcal S$ at the time moment $t \geq 0$, and denote by
$c(t) = m(t)/V$ the number-density of the $X$ particles and by
$c_A$ that of $A$ particles which is kept strongly constant by
suitable reservoir. According to the kinetic law of mass action we
can write
\begin{equation} \label{z1}
\frac{dc(t)}{dt} = k_A c_A\;c(t) - k_{X}\;c^{2}(t) - k_{d}\;c(t) =
k_{X}\;c(t) [c_R - c(t)],
\end{equation}
where
\begin{equation} \label{z2}
c_R = \frac{k_A c_A - k_{d}}{k_{X}}
\end{equation}
is the \textit{ critical parameter} of the reaction. Taking into
account the initial condition $c(0)=c_{X}$ we obtain immediately
the solution of (\ref{z1}) in the form:
\begin{equation} \label{z3}
c(t) = \frac{c_{X} c_R \exp\{c_R k_{X}\; t\}} {c_R +
c_{X}(\exp\{c_R k_{X}\; t\} - 1)}, \;\;\;\;\;\; \text{if} \;c_R
\neq 0, \;\;\;\;\;\; \text{and}\;\;\;\;\;\; c(t) = \frac{c_{X}}{1
+ c_{X} k_{X}\;t}, \;\;\;\;\;\; \text{if} \;c_R = 0.
\end{equation}
Clearly we have two stationary solutions, namely
\begin{equation} \label{z4}
c_{st}^{(1)} = c_R, \;\;\;\;\;\; \text{if} \;\;\;\;\;\; c_R
> 0 \;\;\;\;\;\; \text{and} \;\;\;\;\;c_{st}^{(2)} = 0,
\;\;\;\;\;\; \text{if} \;\;\;\;\;\; c_R \leq 0.
\end{equation}
It is easy to show that the solution of $c_{st}^{(1)}$ is stable,
if $ c_A > k_d/k_A = c_{bf}$,  i.e. if $c_R > 0$, while
$c_{st}^{(2)}$ is stable, if $c_A \leq k_d/k_A = c_{bf}$, i.e. if
$c_R \leq 0$. Let $\delta c$ a small disturbance in $c$. It
follows from Eq. (\ref{z1}) that
\[ \frac{d \delta c}{dt} \approx k_{X}\;[c_R - 2
c(t)]\;\delta c, \] and so, we see immediately, if $c(t)
\rightarrow c_{st}^{(1)}=c_R$, then $\delta c \approx \exp\{-k_{X}
c_R t\}$, i.e. $c_{st}^{(1)}$ is stable, when $c_R
> 0$. Similarly, if $c(t) \rightarrow c_{st}^{(2)}=0$, then
$\delta c \approx \exp\{k_{X} c_R t\}$, i.e. $c_{st}^{(2)}$ is
stable, when $c_R < 0$.

The stationary density of $X$ particles versus $c_A$ can be seen
in FIG. 1. It is clear that $c_A = c_{bf}$ is a "bifurcation
point", since if $c_A > c_{bf}$ then there are two stationary
solutions but among them only one, namely the solution
$c_{st}^{(1)}$ is stable. By decreasing the density $c_A$ \textit{
adiabatically} below the critical value $c_{bf}$, the
autocatalytic particles $X$ are dying out completely, and it is
impossible to start again the process by increasing the density
$c_A$ above the critical $c_{bf}$.

\begin{center}
\setlength{\unitlength}{1cm}
\begin{picture}(5,6)\thinlines
\put(0,0){\vector(0,1){5}} \put(0,0){\vector(1,0){6}} \thicklines
\put(2,0){\line(3,4){3}} \put(0,0){\line(1,0){2}}
\put(0,5.5){\makebox(0,0)[t]{$c_{st}$}}
\put(6.5,0){\makebox(0,0)[t]{$c_A$}}
\put(2,-0.5){\makebox(0,0){$c_{bf}$}} \thinlines \put(1,
2.5){\vector(0,-1){2.5}} \put(5, 2.5){\vector(-1,0){1.1}}
\thicklines \put(1.1,3){\makebox(0,0)[t]{$c_{st}^{(2)}$}}
\put(5.4,2.8){\makebox(0,0)[t]{$c_{st}^{(1)}$}}
\end{picture}
\end{center}

\vspace{0.3cm}

\noindent FIG.$\;\text{1}$: {\footnotesize Dependence of the
stationary number-density of $X$ particles on the number-density
$c_A$. The thick lines refer to stable stationary values.}

\vspace{0.2cm}

In the next subsection we will analyze the stochastic model of
reversible reactions (\ref{a1}). It will be shown that in the
stochastic model the long time behavior of the number of
autocatalytic particles is completely different from that we
obtained by using the rate equation (\ref{z1}).

It is to mention that interesting and seemingly convincing
speculations were published~\cite{parmon02,segre98,stadler93}
about the possibility of "natural selection" based on
autocatalytic reactions in sets of prebiotic organic molecules and
about the "origin of life" beginning with a set of simple organic
molecules capable of self-reproduction. The essence of these
speculations can be summarized as follows: let us consider
$\ell>1$ different and independent autocatalytic particles, and
denote by $c_{bf}^{(1)} < c_{bf}^{(2)} < \cdots < c_{bf}^{(\ell)}$
the corresponding bifurcation points. If $c_A > c_{bf}^{(\ell)}$,
then the stationary density of all autocatalytic particles is
larger than zero. When the density $c_A$ decreases \textit{
adiabatically} to a value lying in the interval $(c_{bf}^{(j-1)},
c_{bf}^{(j)}), \;\; j < \ell$, then the autocatalytic particles
${X}_{j}, {X}_{j+1}, \ldots, {X}_{\ell}$ disappear from the
system, and by increasing again the density of $A$ particles above
$c_{bf}^{(\ell)}$, there is no possibility to recreate those
particles which were lost. In this way some form of selection can
be realized. When we take into account the stochastic nature of
the autocatalytic reaction, then we will see that speculations of
this kind cannot be accepted.

\subsection{Forward equation}

Let the random function $\xi(t)$ be the number of autocatalytic
particles $X$ at the time moment $t \geq 0$. The task is to
determine the probability
\begin{equation} \label{z5}
{\mathcal P}\{\xi(t)=n \vert \xi(0)=N_{X}\} = p_{n}(t)
\end{equation} of finding exactly $n$ autocatalytic particles
$X$ in the system $\mathcal S$ at time instant $t \geq 0$ provided
that at $t = 0$ the number of ${X}$ particles was $N_{X}$. Assume
that the number of the substrate particles $A$ is kept constant
during the whole process, we can write for $n = 1, 2, \ldots$ the
equation:
\[ p_{n}(t + \Delta t) = p_{n}(t)\left[1 - \alpha N_A
n\;\Delta t - \frac{1}{2}\beta' n(n-1)\;\Delta t - \gamma
n\;\Delta t\right] + \]
\[+ \alpha  N_A (n-1)\;p_{n-1}(t)\;\Delta t +
\left[\frac{1}{2} \beta' n(n+1) + \gamma (n+1)\right]p_{n+1}(t)\;
\Delta t + o(\Delta t), \] and for $n=0$ the equation:
\[p_{0}(t + \Delta t) = p_{0}(t) + \gamma p_{1}(t)\;\Delta t +
o(\Delta t),\] where
\begin{equation} \label{z6}
\alpha = \frac{k_A}{V}, \;\;\;\;\;\; \beta' = \frac{k_{X}}{V}
\;\;\;\;\;\; \text{and} \;\;\;\;\;\; \gamma = k_{d},
\end{equation}
From these equations it follows immediately that
\begin{equation} \label{z7}
\frac{dp_{n}(t)}{dt} = - (\alpha N_A - \beta + \gamma + \beta n)\;
n\;p_{n}(t) + + \alpha N_A(n-1)\;p_{n-1}(t) + (\beta n +
\gamma)(n+1) \;p_{n+1}(t),
\end{equation}  \[ n = 1, 2, \ldots, \]
where $\beta = \beta'/2$. If $n=0$, then
\begin{equation}
\label{z8} \frac{p_{0}(t)}{dt} = \gamma p_{1}(t).
\end{equation}

For the sake of completeness we derive the generating function
equations
\[ g(t, z) = {\bf E}\{z^{\xi(t)}\} = \sum_{n=0}^{\infty}
p_{n}(t)\; z^{n} \] and
\[ g_{exp}(t, y) = {\bf E}\{e^{y\;\xi(t)}\} = \sum_{n=0}^{\infty}
p_{n}(t)\; e^{ny}. \] By using the equations (\ref{z7}) and
(\ref{z8}) it is easy to show that $g(t, z)$ satisfies the partial
differential equation
\begin{equation} \label{z9}
\frac{\partial g(t, z)}{\partial t} = -(1 - z)\left(\alpha N_Az -
\gamma \right) \; \frac{\partial g(t, z)} {\partial z} + \beta
(1-z)z\;\frac{\partial^{2} g(t, z)}{\partial z^{2}},
\end{equation}
while $g_{exp}(t, y)$ the equation
\begin{equation} \label{z10}
\frac{\partial g_{exp}(t, y)}{\partial t} = \left[\alpha N_A
(e^{y}-1) + (\beta-\gamma)(1-e^{-y})\right] \; \frac{\partial
g_{exp}(t, y)}{\partial y} - \beta (1-e^{-y})\; \frac{\partial^{2}
g_{exp}(t, y)}{\partial y^{2}}.
\end{equation}
The initial conditions are $g(0, z) = z^{N_{X}}$ and $g_{exp}(0,
y)=e^{N_{X} y}$, respectively, and in addition it is to note that
$g(t, 1) = g_{exp}(t, 0) = 1$. For many purposes it is convenient
to use the logarithm of the exponential generating function
$g_{exp}(t, y$). Therefore, define the function
\begin{equation} \label{z11}
K(t, y) = \log\;g_{exp}(t, y)
\end{equation}
the derivatives of which at $y=0$, i.e.
\[ \left[\frac{\partial^{j} K(t, y)}{\partial y^{j}}\right]_{y=0} = \kappa_{j}(t),
\;\;\;\;\;\; j = 1, 2, \ldots  \] are the \textit{ cumulants} of
$\xi(t)$. One can immediately obtain the equation
\begin{equation} \label{z12}
\frac{\partial K(t, y)}{\partial t} = \left[\alpha N_A (e^{y}-1) +
(\beta-\gamma)(1-e^{-y})\right]\;\frac{\partial K(t, y)}{\partial
y} - \beta (1-e^{-y})\; \left\{\frac{\partial^{2} K(t,
y)}{\partial y^{2}} + \left[\frac{\partial K(t, y)}{\partial
y}\right]^{2}\right\}.
\end{equation}
Now, the initial condition is  \[ K(0, y) = N_{X}\;y, \;\;\;\;\;\;
\text{while} \;\;\;\;\;\; K(t, 0) = 0. \]

In order to simplify the notations define the vector
\begin{equation} \label{z13}
\vec{\bf p}(t) = \left( \begin{array}{c} p_{0}(t) \\ p_{1}(t) \\
\vdots \\ p_{n}(t) \\ \vdots \end{array} \right)
\end{equation}
and the matrix
\begin{equation} \label{z14}
\mathbf{A} = \begin{pmatrix} -D_{0} & C_{0} & 0 & 0 & 0 & \cdots \\
B_{1} & -D_{1} & C_{1} & 0 & 0 & \cdots \\
0 & B_{2} & -D_{2} & C_{2} & 0 & \cdots \\
0 & 0 & B_{3} & -D_{3} & C_{3} & \cdots \\
\vdots & \vdots & \vdots & \vdots & \vdots & \ddots &
\end{pmatrix},
\end{equation}
where
\begin{eqnarray}
B_{n} & = & a\;(n-1), \;\;\;\;\;\; a = \frac{\alpha N_A}{\beta},  \label{z15} \\
D_{n} & = & n\;(n-1+a+b), \;\;\;\;\;\; b = \frac{\gamma}{\beta},  \label{z16} \\
C_{n} & = & (n+1)\;(n+b), \label{z17}
\end{eqnarray}
and write the Eqs. (\ref{z7}) and (\ref{z8}) in the following
concise form:
\begin{equation} \label{z18}
\frac{d\vec{\mathbf{p}}(u)}{du} = \mathbf{A}\;\vec{\mathbf{p}}(u),
\;\;\;\;\;\; \text{where} \;\;\;\;\;\; u = \beta\;t.
\end{equation}
The formal solution of this equation is
\begin{equation} \label{z19}
\vec{\mathbf{p}}(u) = \exp\{\mathbf{A} u\}\;\vec{\mathbf{p}}(0),
\end{equation}
where the components of $\vec{\mathbf{p}}(0)$ are $p_{n}(0) =
\delta_{n,N_{X}}, \;\; n=0,1, \ldots,$ and the matrix $\mathbf{A}$
is a \textit{ normal Jacobi matrix}. As known, the eigenvalues of
a normal Jacobi matrix are different and real. If
\[ \nu_{0} > \nu_{1} > \cdots > \nu_{k} > \cdots \]
are the eigenvalues of $\mathbf{A}$, then the $n$'th component of
the vector  $\vec{\mathbf{p}}(u)$ can be written in the form
\begin{equation} \label{z20}
p_{n}(u) = \sum_{k=0}^{\infty} w_{nk}\;e^{\nu_{k} u},
\end{equation}
for any ${n \geq 0}$. Since to find the eigenvalues $\nu_{k}, \;
k=0,1,\ldots$ and the coefficients $w_{nk}, \;n,k=0,1, \ldots$ is
not a simple task, we concentrate our efforts only on the
determination of stationary solutions of Eqs. (\ref{z7}) and
(\ref{z8}).

\subsubsection{Stationary probabilities}

In order to show that the limit relations
\begin{equation} \label{z21}
\lim_{u \rightarrow \infty} p_{n}(u) =  w_{n}, \;\;\;\;\;\;
\forall n \geq 0, \;\; n \in {\mathcal Z}_{+}
\end{equation}
exist, we need a \textit{theorem for eigenvalues of the matrix
$\mathbf{A}$} stating that $\nu_{0} = 0$ and $\nu_{k} < 0$ for
every $k \geq 1$. The proof of the theorem can be found in
\textbf{Appendix A}. By using this theorem it follows from Eq.
(\ref{z20}) that
\begin{equation} \label{z22}
\lim_{u \rightarrow \infty} p_{n}(u) =  w_{n0}, \;\;\;\;\;\;
\text{i.e.} \;\;\;\;\;\; w_{n} = w_{n0}, \;\;\;\;\;\;\forall n
\geq 0.
\end{equation}
Since in this case  \[ \lim_{u \rightarrow \infty}
\frac{dp_{n}(u)}{du} = 0, \;\;\;\;\;\; \forall \; n \geq 0, \] we
can write from (\ref{z7}) immediately the stationary equations
\begin{equation} \label{z23}
a\;[n w_{n} - (n-1)w_{n-1}] =  n [(n+1) w_{n+1} - (n-1)w_{n}] +
b\;[(n+1) w_{n+1} - n w_{n}],
\end{equation}
for all $n \geq 1$. Summing up both sides of (\ref{z23}) from
$n=1$ to $n=k$, we obtain
\begin{equation} \label{z24}
a\;k w_{k} = k(k+1)w_{k+1} + b\;(k+1) w_{k+1} - b\;w_{1}.
\end{equation}
Taking into account the Eq. (\ref{z8}) we should write that
\begin{equation} \label{z25}
p_{0}(u) = b\;\int_{0}^{u} p_{1}(u')\;du' = b\;\left[w_{1}\;u +
\sum_{k=1}^{\infty} w_{1k}\;\frac{1 -
e^{-|\nu_{k}|u}}{|\nu_{k}|}\right] = w_{0} + \sum_{k=1}^{\infty}
w_{0k}\; e^{-|\nu_{k}|u},
\end{equation}
and we see that the condition of the stationarity is either $w_{1}
= 0$, or $b = 0$, i.e. $\gamma = 0$.

If $w_{1}=0$, then it follows from (\ref{z24}) that $w_{n}=0$ for
all $n=2,3, \ldots$, and  consequently,
\[ w_{0} = \gamma \;\sum_{k=1}^{\infty} \frac{w_{1k}}{|\nu_{k}|}.
\] Since
\[ \sum_{n=0}^{\infty} w_{n} = w_{0} = 1, \;\;\;\;\;\; \text{obviously} \;\;\;\;\;\;
\gamma\;\sum_{k=1}^{\infty} \frac{w_{1k}}{|\nu_{k}|} = 1. \]

If $b = \gamma/\beta = 0$, i.e. if the autocatalytic particles $X$
do not decay, then $w_{0}=0$ and from (\ref{z24}) one obtains that
\[ w_{k+1} = \frac{a}{k+1}\;w_{k},
\;\;\;\;\;\; \text{i.e.} \;\;\;\;\;\; w_{n} =
\frac{a^{n-1}}{n!}\;w_{1}. \] Taking into account that
$\sum_{n=1}^{\infty} w_{n} = 1$ we can write
\begin{equation} \label{z26}
w_{n} = 0, \;\;\;\;\;\; \text{if} \;\;\;\;\;\; \gamma \neq 0
\;\;\;\;\;\; \text{and} \;\;\;\;\;\; w_{n} =
\frac{a^{n}}{n!}\;\frac{e^{-a}}{1 - e^{-a}},\;\;\;\;\;\; \text{if}
\;\;\;\;\;\; \gamma = 0,
\end{equation}
for $n = 1, 2, \ldots$, while for $n=0$ we have
\begin{equation} \label{z27}
w_{0} = \left\{ \begin{array}{ll} 1, & \text{if $\gamma \neq 0$,} \\
\text{ } & \text{ } \\
0, & \text{if $\gamma = 0$.}
\end{array} \right.
\end{equation}

For the calculation of factorial moments let us introduce the
generating function
\begin{equation} \label{z28}
g_{st}(z) = 1, \;\;\;\;\;\; \text{if} \;\;\;\;\;\; \gamma \neq 0
\;\;\;\;\;\; \text{and} \;\;\;\;\;\; g_{st}(z) = \frac{e^{-a(1-z)}
- e^{-a}}{1 - e^{-a}},\;\;\;\;\;\; \text{if} \;\;\;\;\;\; \gamma =
0.
\end{equation}

\subsubsection{Expectation value and variance}

In order to obtain the equation for the mean value of the number
of $X$ particles we use the generating function equation
(\ref{z10}). Introducing the time parameter $u=\beta t$ we have
\[ \frac{dm_{1}(u)}{du} = m_{1}(u)[a-b+1-m_{1}(u)] -
\beta\;[m_{2}(u)-m_{1}^{2}(u)], \] which clearly shows that the
kinetic law of the mass action is violated, when the variance
$V(u) = m_{2}(u)-m_{1}^{2}(u)$ is not negligible. For the second
moment we get the equation
\[ \frac{dm_{2}(u)}{du}  =  (a+b-1)\;m_{1}(u) +
[2(a-b)+3]\;m_{2}(u) - 2\;m_{3}(u), \] in which appears the third
moment $m_{3}(u)$. There are several methods to find approximate
solution of $m_{1}(t)$, some of them were mentioned already in
\cite{lpal04}. Here, we do not want to discuss the details,
instead we are focussing our attention on the properties of
\textit{the expectation value and variance in stationary state}.

If the decay constant $\gamma$ of $X$ particles is zero, then it
is easy to show from Eq. (\ref{z28}) that the stationary value of
the average number of autocatalytic particles in the system is
equal to
\begin{equation} \label{z30}
m_{1}^{(st)} = \frac{a}{1 - e^{-a}},
\end{equation}
while \textit{ the second factorial moment} is
\begin{equation} \label{z31}
m_{st}^{(2)} = \frac{a^{2}}{1 - e^{-a}} = a\;m_{1}^{(st)},
\end{equation}
consequently, the variance can be written in the form
\begin{equation} \label{z32}
V_{st} = m_{st}^{(2)} + m_{1}^{(st)}[1 - m_{1}^{(st)}] =
m_{1}^{(st)}\left(1 - \frac{a}{e^{a} - 1}\right).
\end{equation}
If $\gamma \neq 0$, then $m_{st}^{(k)} = m_{k}^{(st)} = 0, \;
\forall k \geq 1$.

The first important conclusion drawn from the stochastic model is
that the average number of the autocatalytic particles in
stationary state is different from zero, only when the decay rate
constant $\gamma$ of $X$ particles is zero. It means that \textit{
there is no "bifurcation" point in the dependence of the average
stationary number of $X$ particles on the number of $A$
particles}, therefore speculations mentioned earlier about the
"natural selection" are not supported by the stochastic model.

The second conclusion is connected with \textit{the law of the
mass action} referring to the \textit{ chemical equilibrium}. If
$\gamma=0$, then the reversible reaction $A + X \rightleftharpoons
X + X$ has to lead to an equilibrium state, in which
\begin{equation} \label{z33}
m_{1}^{(st)} = N_A\;\frac{k_A}{k_{X}} =
N_A\;\frac{\alpha}{2\;\beta} = \frac{1}{2}\;a.
\end{equation}
The stochastic model results in an entirely different expression,
namely
\begin{equation} \label{z34}
m_{1}^{(st)} = \frac{a}{1 - e^{-a}}.
\end{equation}
The relative dispersion, i.e.
\[ \frac{V_{st}}{m_{1}^{(st)}} = 1 - \frac{a}{e^{a} -1} \] clearly
shows that the fluctuations of the number of autocatalytic
particles become Poisson-like when the number of substrate
particles is increasing.

\section{Modified stochastic model}

The modification of the stochastic model is very simple. We assume
that the probability of the reverse reaction $X + X
\longrightarrow A + X$ is proportional to the average number of
$X$ particles. Therefore, the probability that a reverse reaction
occurs in the time interval $(t, t+\Delta t)$ is nothing else than
$\beta\;m_{1}(t)\;n\;\Delta t + o(\Delta t)$ provided that the
number of $X$ particles was exactly $n$ at time moment $t$.
Accepting this assumption we can rewrite~\footnote{In this case we
use the notation $\tilde{p}_{n}(t)$ instead of $p_{n}(t)$.} the
equations (\ref{z7}) and (\ref{z8}) in the form:
\[ \frac{d\tilde{p}_{n}(t)}{dt} = - [\alpha N_A + \gamma
+ \beta m_{1}(t)]\; n\;\tilde{p}_{n}(t) + \alpha
N_A(n-1)\;\tilde{p}_{n-1}(t) +  \]
\begin{equation} \label{d1}
+ [\gamma + \beta m_{1}(t)](n+1) \;\tilde{p}_{n+1}(t),
\;\;\;\;\;\; n = 1, 2, \ldots,
\end{equation}
and
\begin{equation}
\label{d2} \frac{\tilde{p}_{0}(t)}{dt} = \gamma \tilde{p}_{1}(t),
\end{equation}
respectively. One can immediately see that the generating function
\[ \tilde{g}_{exp}(t, y) = \sum_{n=0}^{\infty}
\tilde{p}_{n}(t)\;e^{ny} \] satisfies the equation
\begin{equation} \label{d3}
\frac{\partial \tilde{g}_{exp}(t, y)}{\partial t} = \left\{N_A
\alpha (e^{y} - 1) - [\beta m_{1}(t) + \gamma](1 -
e^{-y})\right\}\;\frac{\partial \tilde{g}_{exp}(t, y)}{\partial
y},
\end{equation}
with $\tilde{g}_{exp}(0, y) = e^{yN_X}$ and $\tilde{g}_{exp}(t, 0)
= 1$. In the sequel $N_X=1$. Introducing the notations $a=N_A
\alpha/\beta, \; b=\gamma/\beta$ and $u=\beta t$, it is not
surprising that the first moment $m_{1}(u)$ is the solution of the
equation
\begin{equation} \label{d4}
\frac{dm_{1}(u)}{du} = m_{1}(u)\;[a-b -m_{1}(u)],
\end{equation}
which is exactly the same as (\ref{z1}). If the initial condition
is $m_{u}(0)=1$, then
\begin{equation} \label{d5}
m_{1}(u) = \frac{(a-b)\;e^{(a-b)u}}{a-b-1+e^{(a-b)u}},
\end{equation}
and one can see that
\[ \lim_{u \rightarrow \infty} m_{1}(u) = \left\{ \begin{array}{ll}
a-b, &
\text{if $N_A > \gamma/\alpha$, \hspace{0.3cm} i.e. if $a>b$,}  \\
\text{ } &  \text{ } \\
0, & \text{if $N_A < \gamma/\alpha,$ \hspace{0.3cm} i.e. if
$a<b$.}
\end{array} \right. \]
If $N_A = \gamma/\alpha = N_A^{(bf)}$, i.e. if $a=b$, then one
obtains from (\ref{d5})
\[ m_{1}(u) = \frac{1}{1 + u}.  \]
The value $N_A^{(bf)}$ is the number of $A$ particles that
corresponds to \textit{the bifurcation concentration $c_{bf}$}
introduced in the rate equation model.

However, the modified stochastic model takes into account the
randomness of the reactions, and hence gives a possibility for the
determination of \textit{the variance of the number of $X$
particles} versus time. It can be easily shown that the variance
of $\xi(u)$ is
\begin{equation} \label{d7}
{\bf D}^{2}\{\xi(u)\} = m_{2}(u) - m_{1}^{2}(u) = m_{1}(t) +
m_{1}^{2}(u)\;\left[2a \int_{0}^{u}\frac{du'}{m_{1}(u')} - 1
\right].
\end{equation}
In order to prove this relation we need the equation for the
second moment $m_{2}(u)$, which can be derived from Eq.
(\ref{d3}). Introducing the time parameter $u=\beta t$ we obtain
\[ \frac{dm_{2}(u)}{du} = [a+b+m_{1}(u)]\;m_{1}(u) + 2[a-b-m_{1}(u)]\;m_{2}(u), \]
and this can be rewritten in the form
\[ \frac{dm_{2}(u)}{du} = 2a\;m_{1}(u) -
\frac{dm_{1}(u)}{du} + 2m_{2}(u)\;\frac{d \log m_{1}(u)}{du}. \]
It is an elementary task to show that
\[ m_{2}(u) = m_{1}(u) + 2a\;m_{1}^{2}(u)
\int_{0}^{u}\frac{du'}{m_{1}(u')}, \] and from this we obtain
immediately the variance (\ref{d7}).

Taking into account the expression (\ref{d5}) for $m_{1}(u)$ the
variance of the number of $X$ particles can be written in the
following form
\begin{equation} \label{d8}
\mathbf{D}^{2}(u) = \left\{ \begin{array}{ll} m_{1}(u) +
m_{1}^{2}(u)\left[2\;\psi(t) - 1\right], & \text{if
$N_A \neq N_A^{(bf)}$,} \\
\text{ } & \text{ } \\
\left[b u^{2} + (2b + 1) u \right]/(1 + u)^{2}, & \text{if $N_A =
N_A^{(bf)}$,}
\end{array} \right.
\end{equation}
where
\[ \psi(u) = \frac{au}{a-b}
+ \frac{a(a-b-1)}{(a-b)^{2}}\;\left[1 - e^{-(a-b)u} \right].
\] From (\ref{d7}) we can conclude that
\begin{equation} \label{d9}
\lim_{t \rightarrow \infty} \mathbf{D}^{2}\{\xi(u)\} = \left\{
\begin{array}{ll}
0, & \text{if $N_A < N_A^{(bf)}$, } \\
\text{ } & \text{ } \\
b, & \text{if $N_A = N_A^{(bf)}$, } \\
\text{ } & \text{ } \\
\infty, & \text{if $N_A > N_A^{(bf)}$.}
\end{array} \right.
\end{equation}

\setcounter{figure}{1}

\begin{figure} [ht!]
\protect \centering{
\includegraphics[height=6cm, width=8cm]{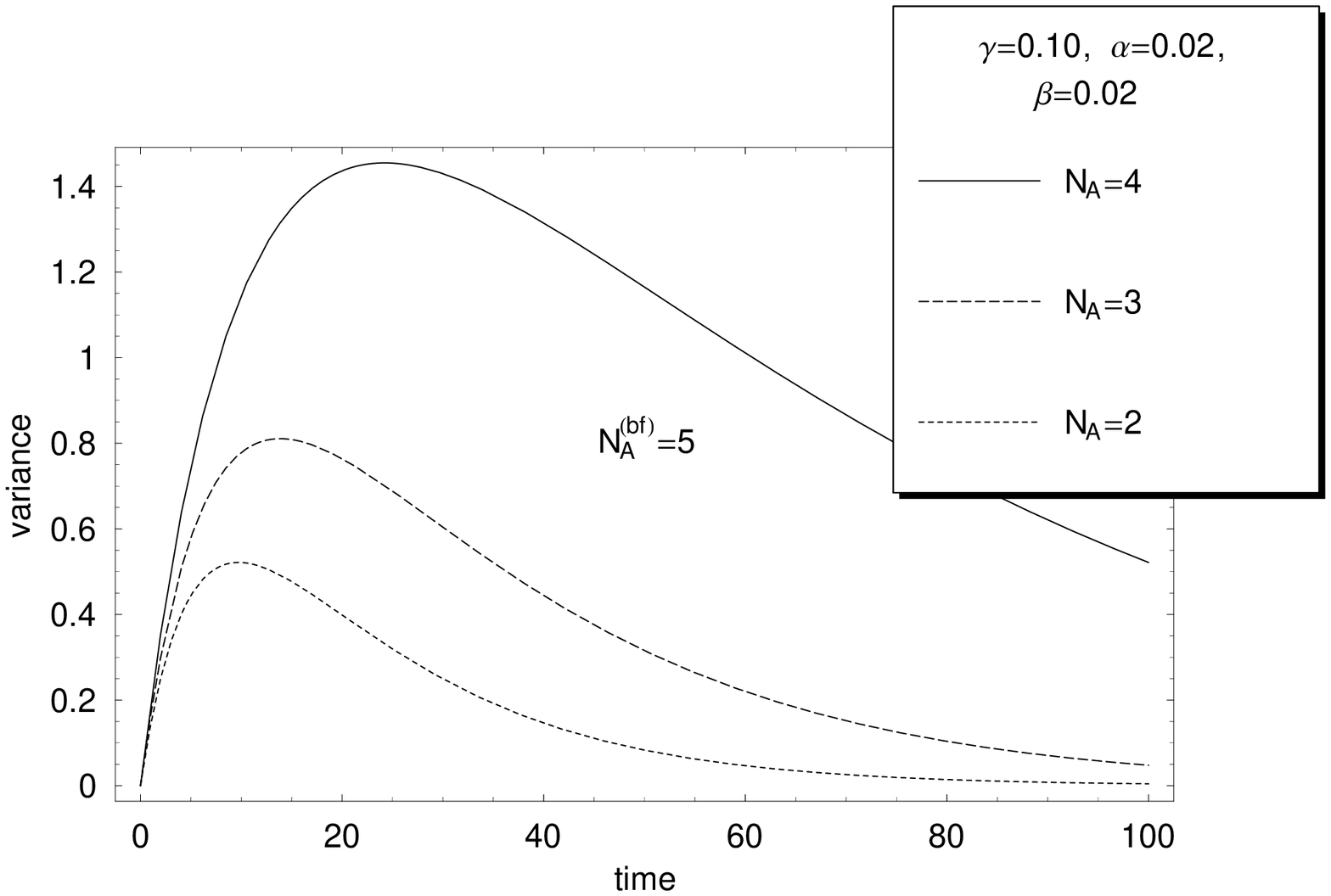}}\protect
\vskip 0.2cm \protect \caption{\label{fig1}{\footnotesize Time
dependence of the variance of the  number of $X$ particles in the
case of $N_A < N_A^{(bf)}$.}}
\end{figure}

\begin{figure} [ht!]
\protect \centering{
\includegraphics[height=6cm, width=8cm]{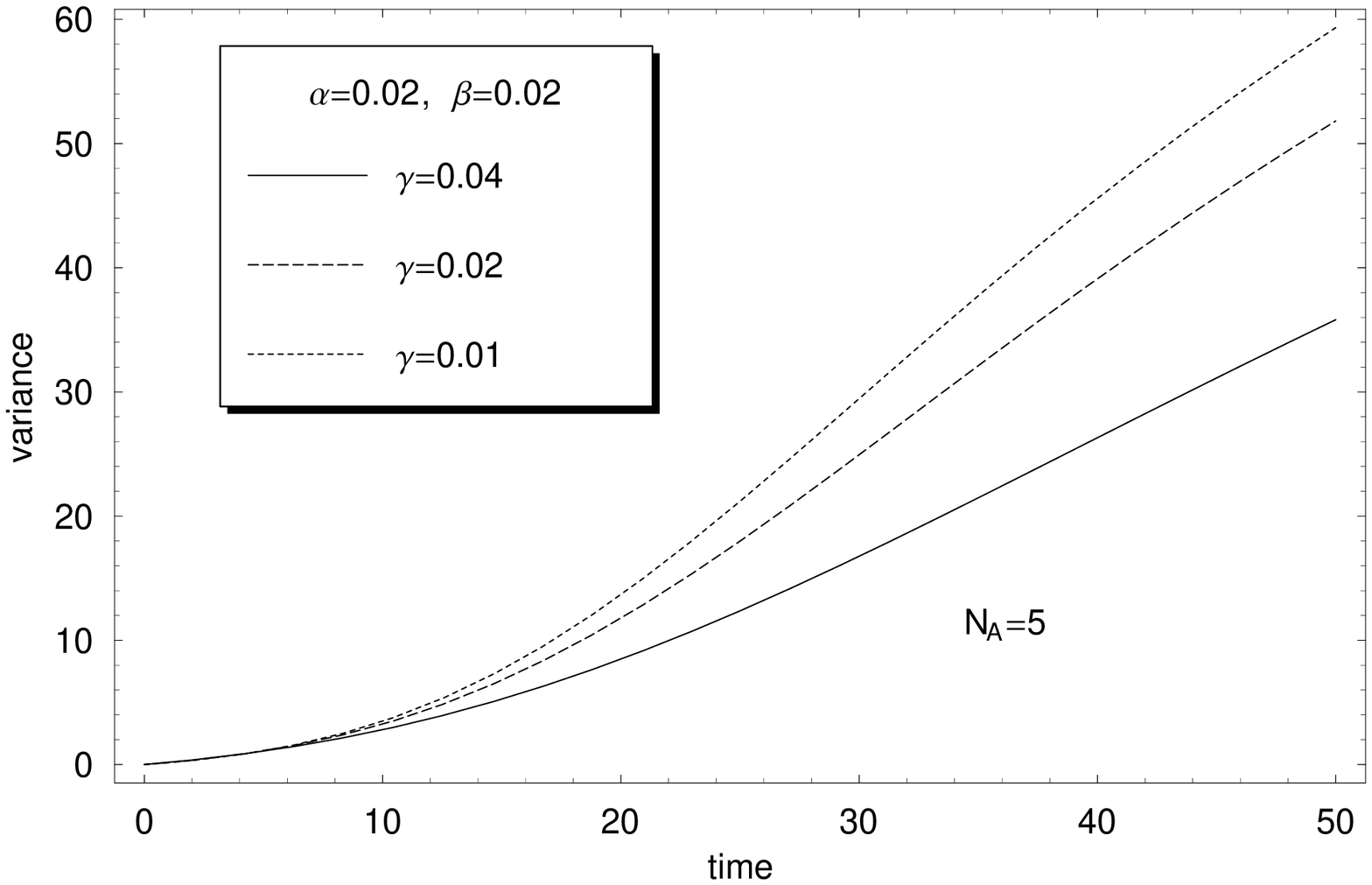}}\protect
\vskip 0.2cm \protect \caption{\label{fig2} {\footnotesize Time
dependence of the variance of the  number of $X$ particles in the
case of $N_A > N_A^{(bf)}$ at different decay rate constants.}}
\end{figure}

In order to have some insight into the nature of the time behavior
of the variance of the number of autocatalytic particles  we have
calculated the variance ${\bf D}^{2}\{\xi(u)\}$ versus time curves
for different values of the number of $A$ particles and for
several decay rate constants $\gamma$. FIG.~\ref{fig1} shows the
time dependence of ${\bf D}^{2}\{\xi(u)\}$ when $N_A < N_A^{(bf)}$
at a fixed value of rate constant. We see that the variance versus
time curves have a maximum and after that they decrease slowly to
zero.

In contrary to this, when $N_A > N_A^{(bf)}$, then the variance of
the number of autocatalytic particles increase mono\-ton\-ously to
infinity. It can be easily shown that
\[ \lim_{u \rightarrow \infty} \frac{\mathbf{D}^{2}\{\xi(u)\}}{u} =
\frac{1}{N_A}\frac{a}{a-b}, \] and so, we can use for large $u$
the asymptotic formula
\[ {\bf D}^{2}\{\xi(u)\} \approx \frac{1}{N_A}\frac{a}{a-b}\;u,
\;\;\;\;\;\;  u >> 1. \] The time dependence of the variance ${\bf
D}^{2}\{\xi(u)\}$ in the case when $N_A > N_A^{(bf)}$ is seen in
FIG.~\ref{fig2} for three decay rate constants. The variance
converges to the infinity linearly with increasing time parameter,
if $N_A > N_A^{(bf)}$, and to a constant value $b$, if $N_A =
N_A^{(bf)}$, so we can say that the fluctuation of the number of
$X$ particles in the stationary state near the "bifurcation" point
alters the possible conclusions based only on the average number
of autocatalytic particles.

\section{Lifetime of the system}

We say that a system is in the state ${\mathcal S}_{j}(t)$, when
$\xi(t)=j$. Obviously, the system is live at time instant $t \geq
0$, when $\xi(t) > 0$. Let us define the probability of transition
${\mathcal S}_{i}(t_0) \rightarrow {\mathcal S}_{j}(t_0+t)$ by
\begin{equation} \label{z35}
{\mathcal P}\{{\mathcal S}_{j}(t_0+t)|{\mathcal S}_{i}(t_0)\} =
p_{i,j}(t_0+t, t_0), \;\;\;\;\;\; \text{where} \;\;\;\;\;\; t \geq
0.
\end{equation}
Since the process is homogenous in time $p_{i,j}(t_0+t, t_0) =
p_{i,j}(t)$, and using Eq. (\ref{z7}) we can write that
\begin{equation} \label{z36}
\frac{dp_{i,j}(t)}{dt} = -(\lambda_{j} + \mu_{j})\;p_{i,j}(t) +
\lambda_{j-1}\;p_{i,j-1}(t) + \mu_{j+1}\;p_{i,j+1}(t),
\;\;\;\;\;\; i,j=0,1, \ldots,
\end{equation}
where
\begin{equation} \label{z37}
\lambda_{j} = j N_{X} \alpha , \;\;\;\;\;\; \mu_{j} = j[\beta
(j-1) + \gamma].
\end{equation}
We see that the first equation is
\begin{equation} \label{z38}
\frac{dp_{i,0}(t)}{dt} = \gamma p_{i,1}(t),
\end{equation}
and the initial condition is given by $p_{i,j}(0) = \delta_{i,j}$.
The random time $\theta_{n}$ due to the transition ${\mathcal
S}_{n}(0) \rightarrow {\mathcal S}_{0}(\theta_{n}),\; n > 0$ is
\textit{ the lifetime of the system} which has been at $t=0$ in
the state ${\mathcal S}_{n}(0)$. It is evident that
\[ {\mathcal P}\{\theta_{n} \geq t\} = {\mathcal
P}\{\xi(t)=0|\xi(0)=n\} = p_{n,0}(t) = H_{n}(t), \] where
$H_{n}(t)$ is the probability that the lifetime $\theta_{n}$ is
not larger than $t$. The moments of the lifetime are given by
\begin{equation} \label{z39}
\mathbf{E}\{\theta_{n}^{k}\} = \int_{0}^{\infty} t^{k}\;dH_{n}(t),
\;\;\;\;\;\; n > 0 \;\;\;\;\;\; \text{and} \;\;\;\;\;\;
k=1,2,\ldots\;\;\;.
\end{equation}

\subsection{Extinction probability}

The extinction of a system of state ${\mathcal S}_{n}(0)$ occurs
when a transition to the state ${\mathcal S}_{0}(t)$ is realized
for any $t \geq 0$. We define the extinction probability by
\[ \lim_{t \rightarrow \infty} H_{n}(t) = L_{n}, \;\;\;\;\;\; \forall
\; n = 1, 2, \ldots, \;\;\;\;\;\; \text{and since} \;\;\;\;\;\;
H_n(\infty) = p_{n,0}(\infty) = w_0 \] in accordance with
(\ref{z27}) we find that
\begin{equation} \label{z40}
L_{n} = \left\{\begin{array}{ll} 1, & \text{if $\gamma \neq 0$,}
\\ \text{ } & \text{ } \\ 0, & \text{if $\gamma = 0$,}
\end{array} \right.
\;\;\;\;\;\; \forall \; n = 1, 2, \ldots \;\;.
\end{equation}
It is a remarkable result stating that a system being in any of
states ${\mathcal S}_n(t_0)$ at a given time instant $t_0$ after
elapsing sufficiently long time $t>>t_0$ will be almost surly
annihilated, if $\gamma \neq 0$, and never dies, i.e. the
extinction probability is zero, if $\gamma=0$.

It is to mention that the statement (\ref{z40}) can be obtained
from a nice lemma by Karlin~\cite{karlin75} which can be
formulated in the following way: introducing the notation
\begin{equation} \label{z41}
\rho_k = \prod_{j=1}^{k} \frac{\mu_j}{\lambda_j},
\end{equation}
where $\mu_j$ and $\lambda_j$ are non-negative real numbers which
are not necessarily equal to (\ref{z37}), it can be stated that if
\begin{equation} \label{z42}
K = \lim_{n \rightarrow \infty} \sum_{k=1}^{n}\;\rho_k = \infty,
\;\;\;\;\;\; \text{then} \;\;\;\;\;\; L_{n} = 1, \;\;\;\;\;\;
\forall \; n = 1, 2, \ldots,
\end{equation}
while if
\begin{equation} \label{43}
K = \lim_{n \rightarrow \infty} \sum_{k=1}^{n}\;\rho_k < \infty,
\;\;\;\;\;\; \text{then} \;\;\;\;\;\; L_{n} =
\frac{\sum_{k=n}^{\infty} \rho_{k}}{1 + \sum_{k=1}^{\infty}
\rho_{k}} < 1,\;\;\;\;\;\; \forall \; n = 1, 2, \ldots \;\;.
\end{equation}
By using the expressions (\ref{z37}) we see immediately that
\[ \rho_{k} = \prod_{j=1}^{k}\frac{\beta (j-1) +
\gamma}{N_A \alpha} = \left(\frac{\beta}{N_A
\alpha}\right)^{k}\;\frac{\Gamma(\gamma/\beta+j)}
{\Gamma(\gamma/\beta)}, \] hence we can conclude that
\[ K = \left\{\begin{array}{ll} \infty, & \text{if $\gamma \neq 0$,}
\\ \text{ } & \text{ } \\ 0, & \text{if $\gamma = 0$,}
\end{array} \right.  \]
and so, applying the lemma we prove the statement (\ref{z40}).

\subsection{Average lifetime of the system}

To determine the transition probability $p_{n,0}(t)$, i.e. the
probability $H_{n}(t)$ is not an easy problem. Instead, we show
how to calculate the average lifetime $\mathbf{E}\{\theta_{n}\} =
\tau_{n}$. Let us define the parameters
\begin{equation} \label{z44}
\delta_{1} =  \frac{1}{\mu_{1}}, \;\;\;\;\;\; \text{and}
\;\;\;\;\;\;  \delta_{k} = \frac{\lambda_{1} \cdots
\lambda_{k-1}}{\mu_{1} \cdots \mu_{k}}, \;\;\;\;\;\; \text{if
$k>1$},
\end{equation}
and formulate the following statement called Karlin's theorem
\cite{karlin75}. \textit{ If $\;\sum_{k=1}^{\infty} \delta_{k} <
\infty$, then the average lifetime $\tau_{n}$ of a system
containing $n$ autocatalytic particles is given by
\begin{equation} \label{z45}
\tau_{n} = \sum_{k=0}^{\infty} \delta_{k} + \sum_{j=1}^{n-1}
\rho_{j}\;\sum_{k=j+1}^{\infty} \delta_{k},
\end{equation}
where $\rho_{j}$ is defined by (\ref{z41}), and in contrary, if
$\;\sum_{k=1}^{\infty} \delta_{k} = \infty$, then  $\tau_{n} =
\infty \;\; \forall\; n \geq 1.$} The proof of this statement can
be found in \textbf{Appendix B}.

Now, by using this statement we would like to calculate the
average lifetime of a system which is in the state ${\mathcal
S}_{n}(t)$ at the moment $t$. Introducing the notations
\begin{equation} \label{z46}
a = \frac{N_A \alpha}{\beta} \;\;\;\;\;\; \text{and} \;\;\;\;\;\;
b = \frac{\gamma}{\beta},
\end{equation}
and by using the expressions (\ref{z37}) we can write
\begin{equation} \label{z47}
\delta_{j} = \frac{1}{\gamma}\left( \delta_{j,1} + (1 -
\delta_{j,1})\;\frac{a^{j-1}}{j\;(b+1) \cdots (b+j-1)}\right),
\end{equation}
where $\delta_{j,1}$ is the Kronecker-symbol. Define the sum
\begin{equation} \label{z48}
d_{n} = \sum_{j=1}^{n} \delta_{j} = \frac{1}{\gamma}\left(1 +
\sum_{j=2}^{n} \frac{(j-1)!}{(b+1) \cdots
(b+j-1)}\;\frac{a^{j-1}}{j!}\right) =
\frac{\Gamma(b+1)}{\gamma}\;\sum_{j=1}^{n}
\frac{a^{j-1}}{j\;\Gamma(b+j)},
\end{equation}
for $n \geq 1$. If $\gamma \neq 0$, then one can see immediately
that
\[  \lim_{n \rightarrow \infty} d_{n} = \sum_{j=1}^{\infty} \delta_{j} <
\infty, \]  i.e. the formula (\ref{z45}) should be used for the
calculation of the average lifetime $\tau_{n}$.

First, determine $\tau_{1}$. It follows from Eq. (\ref{z45}) that
\begin{equation} \label{z49}
\tau_{1} = \sum_{j=1}^{\infty} \delta_{j} =
\frac{\Gamma(b+1)}{\gamma}\;\sum_{j=1}^{\infty}\;
\frac{a^{j-1}}{j\;\Gamma(b+j)} = \frac{1}{a\;\gamma}\;\int_{0}^{a}
\Phi(1, b+1; u)\;du,
\end{equation}
where $\Phi(1, b+1; u)$ is the confluent hypergeometric function.
The next step is the calculation of the expression
\[ s_{j}(a, b) = \rho_{j}\;\sum_{k=j+1}^{\infty} \delta_{k}, \]
which can be rewritten into the form:
\[ s_{j}(a, b) = \frac{1}{a^{j}}\;\prod_{i=1}^{j}(b+i-1)\;
\sum_{k=j+1}^{\infty} \delta_{k}.\] After some elementary algebra
we obtain
\[ s_{j}(a, b) =
\frac{1}{\beta}\;\Gamma(b+j)\;
\sum_{k=j+1}^{\infty}\frac{a^{k-j-1}}{k\;\Gamma(b+k)}, \] and
finally we have
\begin{equation} \label{z50}
\tau_{n} = \tau_{1} + \frac{1}{\beta}\;\sum_{j=1}^{n-1}
\Gamma(b+j)\;
\sum_{k=j+1}^{\infty}\frac{a^{k-j-1}}{k\;\Gamma(b+k)}.
\end{equation}
Introducing a new index $\ell = k - j$ we have
\[ s_{j}(a, b) = \frac{1}{\beta}\;\Gamma(b+j)
\sum_{\ell=0}^{\infty}\frac{a^{\ell}}
{(j+\ell+1)\;\Gamma(b+j+\ell+1)}, \] which can be transformed into
the expression
\[ s_{j}(a, b) =  \frac{1}{\beta}\;\frac{1}{a^{j+1}}\;\int_{0}^{a}
v^{j-1}\;\sum_{\ell=0}^{\infty}\frac{(\ell+1)!\;\Gamma(b+j)}
{\Gamma(b+j+\ell+1)}\;\frac{v^{\ell+1}}{(\ell+1)!}\;dv. \]  By
using the identity
\[ \sum_{\ell=0}^{\infty}\frac{(\ell+1)!\;\Gamma(b+j)}
{\Gamma(b+j+\ell+1)}\;\frac{v^{\ell+1}}{(\ell+1)!} = \Phi(1, b+j;
v) - 1, \] $s_{j}(a, b)$ takes a new form, namely
\[ s_{j}(a, b) = \frac{1}{\beta}\; \frac{1}{a^{j+1}}\;\int_{0}^{a}
v^{j-1}\;\left[\Phi(1, b+j; v) - 1\right]\;dv. \] Taking into
account this formula the expression (\ref{z50}) can be rewritten
in the form
\begin{equation} \label{z51}
\tau_{n} = \tau_{1} + \frac{1}{\beta}\;\sum_{j=1}^{n-1}
\frac{1}{a^{j+1}}\;\int_{0}^{a} v^{j-1}\left[\Phi(1, b+j; v) - 1
\right]\;dv,
\end{equation}
which is convenient for numerical calculations. From this equation
we see that \[ \tau_{1} < \tau_{2} < \cdots < \tau_{n} < \cdots
\;, \] and we can prove the limit relation
\begin{equation} \label{z52}
\lim_{n \rightarrow \infty} \beta\;\tau_{n} < \infty.
\end{equation}

\begin{figure} [ht!]
\protect \centering{
\includegraphics[height=6cm, width=9cm]{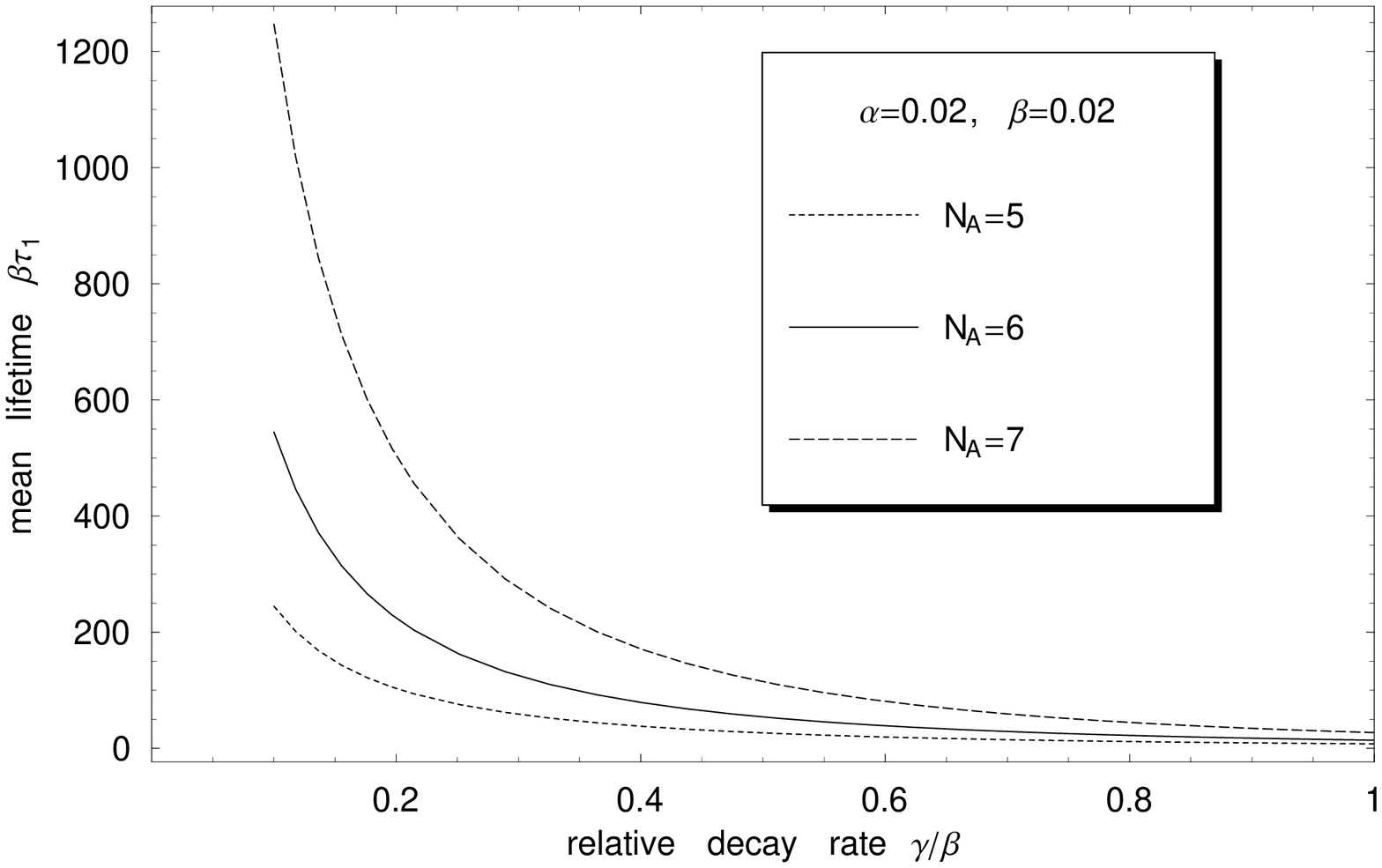}}\protect
\vskip 0.2cm \protect \caption{\label{fig4} {\footnotesize The
mean value of the system lifetime vs. $\gamma/\beta$ at $N_A=5, 6,
7$, provided that the initial state of the system was ${\mathcal
S}_{1}$.}}
\end{figure}

It is \textit{necessary and sufficient} to show that
\[ \beta \;\lim_{n \rightarrow \infty} \sum_{j=1}^{n-1} s_{j}(a,
b) < \infty. \] Since $\Phi(1, b+j; v) - 1$ is a non-negative,
monotonously increasing function of $v \geq 0$ we can write
immediately that
\[ \beta s_{j}(a, b) \leq \frac{\Phi(1, b+j; a) -
1}{a^{j+1}}\; \int_{0}^{a} v^{j-1}\;dv = \frac{\Phi(1, b+j; a) -
1}{a\;j}, \] and if $j \geq 1$, then
\[ \Phi(1, b+j; a) - 1 =
\sum_{k=1}^{\infty}\frac{a^{k}}{(b+j)(b+j+1) \cdots (b+j+k-1)}
\leq  \sum_{k=1}^{\infty} \frac{a^{k}}{j\;(j+1)(j+2) \cdots
(j+k-1)} \leq \frac{a}{j}\;e^{a}, \] consequently, we obtain the
inequality
\[ \beta \;\lim_{n \rightarrow \infty} \sum_{j=1}^{n-1} s_{j}(a,
b) < e^{a}\;\sum_{j=1}^{\infty} \frac{1}{j^{2}} =
\frac{\pi^{2}}{6}\;e^{a} < \infty, \] and this proves the
statement (\ref{z52}).

We calculated how the mean value $\tau_{1}$ depends on the ratio
$\gamma/\beta$ at three different values of the number of $A$
particles. The results are shown in FIG.~\ref{fig4}. We can see
that $\beta \tau_{1}$ decreases rapidly with increasing
$\gamma/\beta$, and if the values $\gamma/\beta$ are smaller than
$0.4$, then we can observe that the larger is the number $N_A$ in
the system the longer is the average lifetime $\beta \tau_{1}$.

\begin{figure} [ht!]
\protect \centering{
\includegraphics[height=6cm, width=9cm]{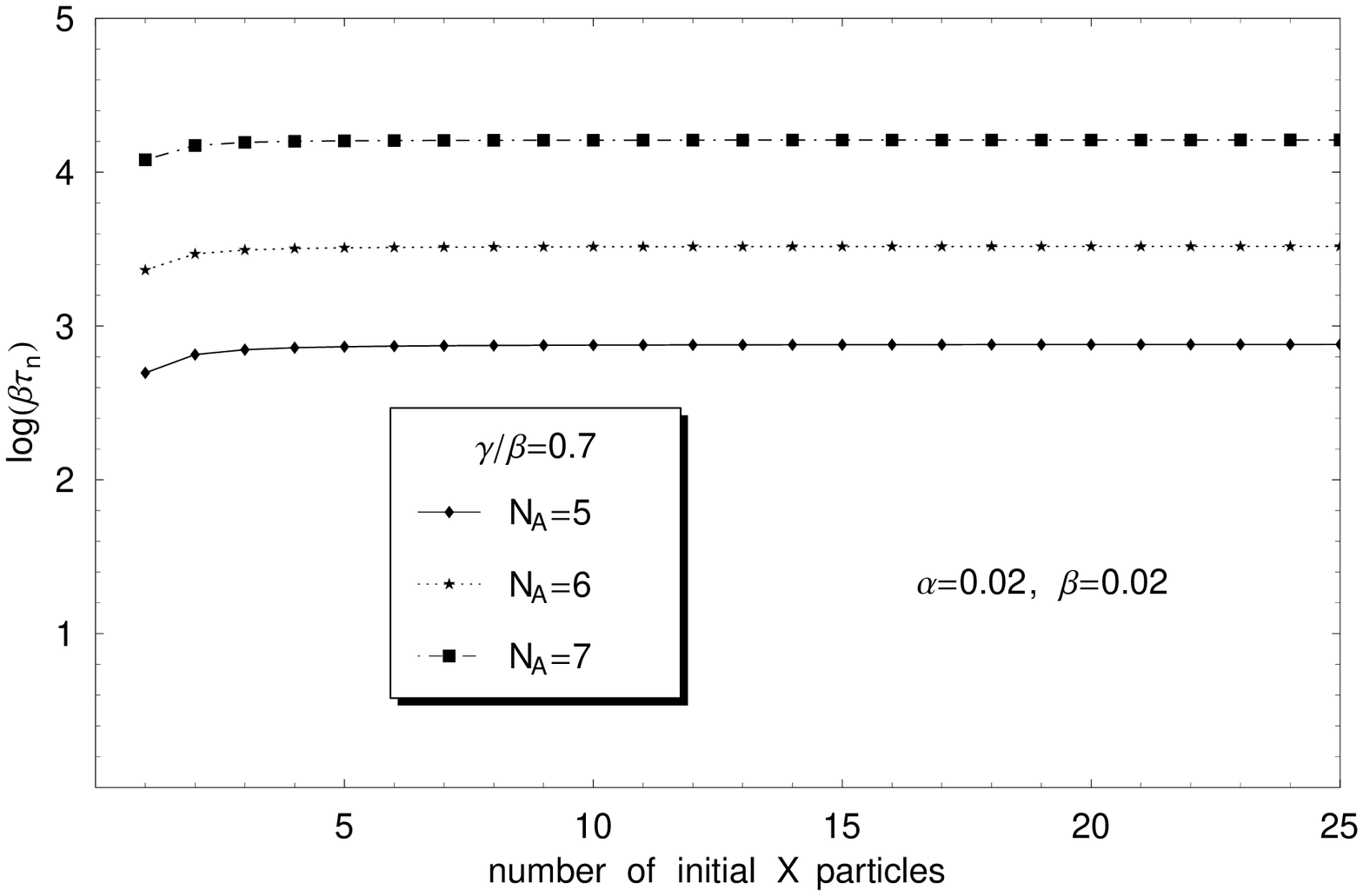}}\protect
\vskip 0.2cm \protect \caption{\label{fig5} {\footnotesize
Dependence of the logarithm of the mean value of the system
lifetime on the number of $X$ particles to be found in the system
at that time moment which the lifetime is counted from.}}
\end{figure}

That FIG.~\ref{fig5} shows is rather surprising. The mean value of
the system lifetime does not depend practically on the number of
$X$ particles to be found in the system at that time moment which
the lifetime is counted from. We can state that the average
lifetime of systems controlled by reactions (\ref{a1}) is already
determined by several ${X}$ particles, and even a large increase
of the number of ${X}$ particles does not effect significantly on
the system lifetime.
\begin{figure} [ht!]
\protect \centering{
\includegraphics[height=6cm, width=9cm]{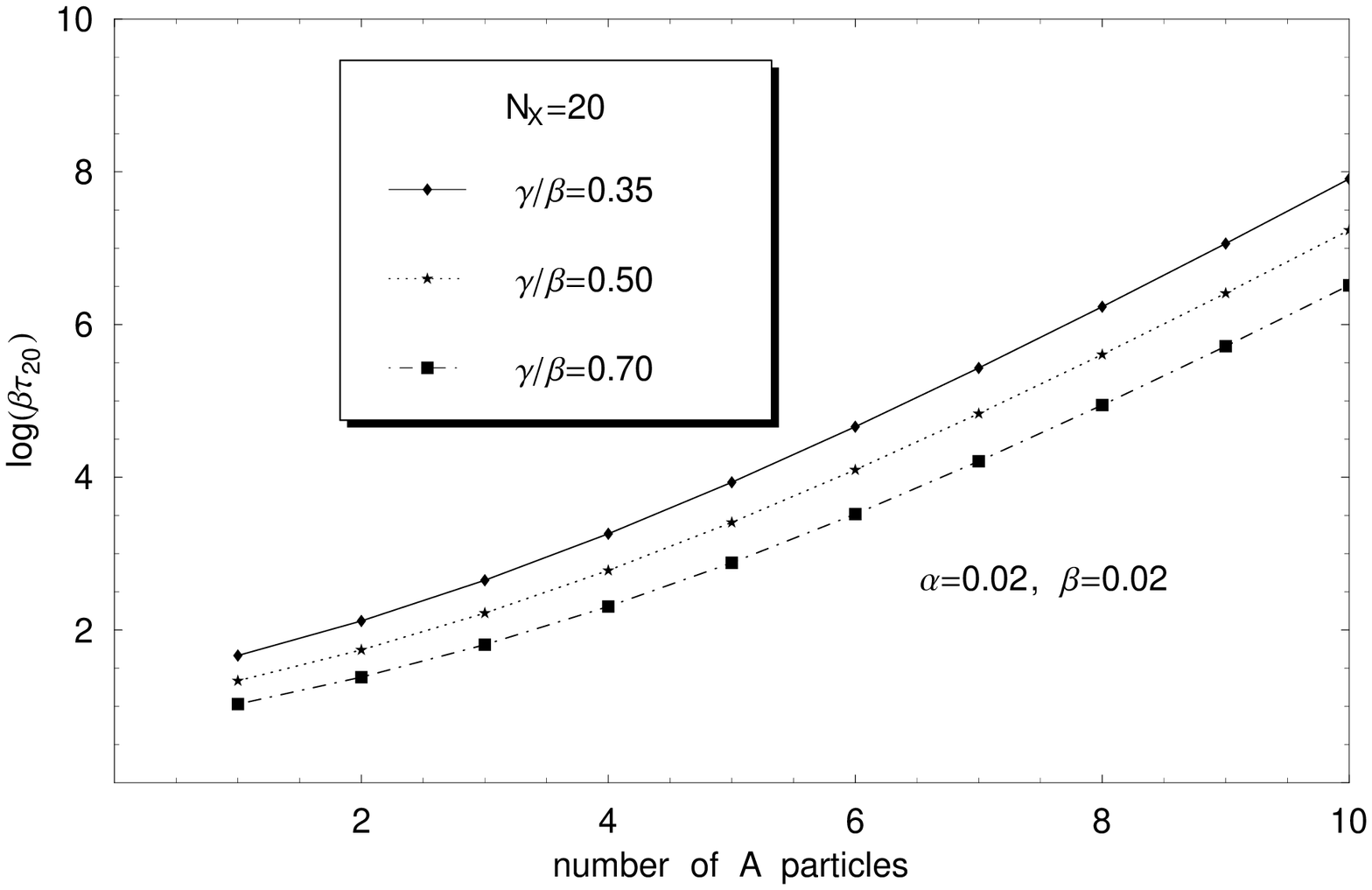}}\protect
\vskip 0.2cm \protect \caption{\label{fig6} {\footnotesize
Dependence of the logarithm of the mean value $\beta\;\tau_{20}$
on the number of $A$ particles at three different values of
$\gamma/\beta$.}}
\end{figure}
On this basis imagine an "organism" which consists of ${X}$
particles capable of self-reproduction and self-annihilation.
Assume that the organism becomes dead if it loses the last ${X}$
particle. One might think that the greater is the number of ${X}$
particles in the organism the larger is irs average lifetime.
Contrary to this conviction, an organism containing, let us say
$500$ particles hardly lives longer than that which contains only
$10$ particles. By using the values $N_A = 7, \; \beta = 0.02$ and
$\gamma = 0.014$ we obtain that $\beta \tau_{10} \approx 67.21$
and $\beta \tau_{500} \approx 67.351$. It is rather surprising
that the increase is only $0.21\%$.

In FIG.~\ref{fig6} we see the dependence of the logarithm of the
mean value $\beta\;\tau_{20}$ on the number of $A$ particles at
three different $\gamma/\beta$. One can observe that the increase
of $N_A$ results in a rather large lengthening of the average
lifetime, i.e. the effect of the substrate particles on the
process is much stronger than that of the autocatalytic particles.

\section{Conclusions}

We assumed the distribution of reacting particles in the system
volume to be uniform and introduced the notion of \textit{ the
point model of reaction kinetics}. In this model the probability
of a reaction between two particles per unit time is evidently
proportional to the product of their actual numbers. By using this
assumption we constructed a stochastic model for systems
controlled by the reactions $A + X \rightleftharpoons X + X, \;\;
X \rightarrow B$ provided that the number of $A$ particles is kept
constant by a suitable reservoir, and the end-product particles
$B$ do not take part in the reaction.

We have shown that the stochastic model results in an equation for
the expectation value $m_{1}(t)$ of autocatalytic particles $X$
which differs strongly from the kinetic rate equation. Further, we
found that if the decay constant $\gamma$ of the particles $X$ is
not zero, then \textit{the stochastic description}, in contrary to
the rate equation description, \textit{brings about only one
stationary state with probability $1$}, and it is the zero-state
${\mathcal S}_{0}$ . It has been also proven that the probability
of a nonzero stationary state is larger than zero, if and only if
the decay rate constant is equal to zero. Consequently, the
average number of $X$ particles in the stationary state is larger
than zero, if only $\gamma=0$. However, one has to underline that
this average number is completely different from that which
corresponds to the law of the mass action of reversible chemical
reactions.

We paid a special attention on the random behavior of \textit{the
system lifetime}, and derived an exact formula for the average
lifetime. It has been shown that the mean value of the system
lifetime does not depend practically on the number of $X$
particles to be found in the system at the time instant which the
lifetime is counted from. For example, the lifetime of a system
having $500$ $X$ particles at the beginning is larger only by
$0.2$\% than the lifetime of a system containing $10$ $X$
particles at the time moment $t=0$.

\appendix

\section{Eigenvalues of the matrix $\mathbf{A}$}

As mentioned already, the eigenvalues of a normal Jacobi matrix
are real and different. We would like to prove the following
theorem:

\vspace{0.2cm}

\textbf{Theorem.} \textit{ The $\nu =0$ is an eigenvalue of the
normal Jacobi matrix ${\bf A}$ defined by (\ref{z14}) and the
other nonzero eigenvalues are all negative.}

\textbf{Proof.} Denote by $a_{ij}, \;\; i,j \in {\mathcal
Z}_{+}$~\footnote{${\mathcal Z}_{+}$ is the set of nonnegative
integers.} the elements of matrix $\mathbf{A}$. We see immediately
that
\begin{equation} \label{c1}
\sum_{i} a_{ij} = C_{j-1} - D_{j} + B_{j+1} = 0,
\end{equation}
i.e. the sum of elements of any column of the matrix $\mathbf{A}$
is equal to zero. Strictly speaking, the theorem itself is a
straightforward consequence of this property. If $\nu$ an
eigenvalue, then
\[ \sum_{j} a_{ij} x_{j} = \nu x_{i}, \]
and $x_{0}, x_{1}, \ldots, x_{i}, \ldots $ define a nonzero
eigenvector
\[ \vec{x} = \left(\begin{array}{c}
x_{0} \\ x_{1} \\ \vdots \\ x_{j} \\ \vdots \end{array} \right).
\]
Let $\vec{y}$ be an arbitrary nonzero vector and form the
following expression:
\[ \sum_{i}\;\sum_{j} a_{ij} x_{j} y_{i} = \nu \sum_{i} x_{i} y_{i} =
\nu \sum_{j} x_{j} y_{j}, \] which can be rewritten in the form
\[ \sum_{j} x_{j} \left(\sum_{i} a_{ij} y_{i} - \nu y_{j}\right)
= 0. \] This equality can be valid only if
\begin{equation} \label{c2}
\sum_{i} a_{i,j} y_{i} = \nu y_{j},
\end{equation}
and because $\vec{y}$ is an arbitrary nonzero vector one can chose
its components to be equal to unity. Then taking into account the
property (\ref{c1}) one has
\[ \nu = \sum_{i} a_{i,j} = 0,  \]
i.e. $\nu = 0$ is indeed an eigenvalue.

Now, we show if $\nu \neq 0$, then $\nu < 0$. Since $a_{jj} \leq
0$, and $a_{ij} \geq 0$, if $i \neq j$, it follows from (\ref{c2})
that
\begin{equation} \label{c3}
\sum_{i \neq j} a_{ij} y_{i} = \left(|a_{jj}| + \nu \right)y_{j}.
\end{equation}
As $\vec{y}$ is an arbitrary nonzero vector, let us chose it so
that
\[ |y_{j}| = \max_{i} |y_{i}| = q > 0, \;\;\;\;\;\; \text{and since}
\;\;\;\;\;\; \sum_{i \neq j} a_{ij} = |a_{jj}|, \] we obtain the
inequality
\[ \sum_{i \neq j} a_{ij} y_{i} \leq \sum_{i \neq j} a_{ij} q = q
|a_{jj}|. \] Taking into account the relation (\ref{c3}) we have
\[ q |a_{jj}| \geq \left(|a_{jj}| + \nu
\right) q \] which can be valid only if $\nu < 0$. Q.E.D.

\section{Karlin's theorem}

\textbf{Theorem.} \textit{If $\;\sum_{k=1}^{\infty} \delta_{k} <
\infty$, then the average lifetime $\tau_{n}$ of a system
containing $n$ autocatalytic particles is given by
\begin{equation} \label{e0}
\tau_{n} = \sum_{k=0}^{\infty} \delta_{k} + \sum_{j=1}^{n-1}
\rho_{j}\;\sum_{k=j+1}^{\infty} \delta_{k},
\end{equation}
where $\rho_{j}$ is defined by (\ref{z41}), and in contrary, if
$\;\sum_{k=1}^{\infty} \delta_{k} = \infty$, then  $\tau_{n} =
\infty \;\; \forall\; n \geq 1.$}

\textbf{Proof.} Assume the system to be in the state ${\mathcal
S}_{n}$ at a given time instant $t$ and suppose that the first
reaction after $t$ occurs at a random time moment
$t+\vartheta_{n}$. This reaction can result in a transition to
either the state ${\mathcal S}_{n-1}$ or ${\mathcal S}_{n+1}$ with
probabilities
\[ \frac{\mu_{n}}{\lambda_{n} + \mu_{n}} \;\;\;\;\;\;\; \text{and}
\;\;\;\;\;\; \frac{\lambda_{n}}{\lambda_{n} + \mu_{n}}, \]
respectively. Since $t$ is arbitrary, the equation
\begin{equation} \label{e1}
\theta_{n} = \vartheta_{n} + \frac{\mu_{n}}{\lambda_{n} +
\mu_{n}}\;\theta_{n-1} + \frac{\lambda_{n}}{\lambda_{n} +
\mu_{n}}\; \theta_{n+1}
\end{equation}
is valid with probability $1$. Taking into account that
\[ {\mathcal P}\{\vartheta_{n} \geq t\} =  1 - \exp\{-(\lambda_{m}
+ \mu_{n})t\}, \] one obtains from (\ref{e1}) the recursion
\begin{equation} \label{e2}
\tau_{n} = \frac{1}{\lambda_{n} + \mu_{n}} +
\frac{\mu_{n}}{\lambda_{n} + \mu_{n}}\;\tau_{n-1} +
\frac{\lambda_{n}}{\lambda_{n} + \mu_{n}}\;\tau_{n+1},
\end{equation}
where $\tau_{0}=0$. Introducing the difference $\omega_{n} =
\tau_{n} - \tau_{n+1}$ after simple rearrangements we obtain
\begin{equation} \label{e3}
\omega_{n} = \frac{1}{\lambda_{n}} +
\frac{\mu_{n}}{\lambda_{n}}\;\omega_{n-1},
\end{equation}
the solution of which can be written in the form:
\[ \omega_{n} = \frac{1}{\lambda_{n}} +
\sum_{i=1}^{n-1}\frac{1}{\lambda_{i}}\;\prod_{j=i+1}^{n}
\frac{\mu_{j}}{\lambda_{j}} +
\omega_{0}\;\prod_{j=1}^{n}\frac{\mu_{j}}{\lambda_{j}}. \] By
using the notation
\[ \delta_{i} = \left\{\begin{array}{ll} \frac{1}{\mu_{1}}, &
\text{if $i=1$,} \\ \text{ } & \text{ } \\
\frac{\lambda_{1} \cdots \lambda_{i-1}}{\mu_{1} \cdots \mu_{i}}, &
\text{if $i>1$.  } \end{array} \right. \] and taking into account
the identity
\[ \left(\sum_{i=1}^{n} \delta_{i}\right)\;
\prod_{j=1}^{n}\frac{\mu_{j}}{\lambda_{j}} = \frac{1}{\lambda_{n}}
+ \sum_{i=1}^{n-1} \frac{1}{\lambda_{i}} \;\prod_{j=i+1}^{n}
\frac{\mu_{j}}{\lambda_{j}}, \] and the relation $\omega_{0} =
\tau_{1}$, we have
\begin{equation} \label{e4}
(\tau_{n} - \tau_{n+1})\;\prod_{j=1}^{n}
\frac{\lambda_{j}}{\mu_{j}} = \sum_{i=1}^{n} \delta_{i} -
\tau_{1}.
\end{equation}

If $\lim_{n \rightarrow \infty} \sum_{i=1}^{n} \delta_{i} =
{\mathcal D} = \infty$, then $\tau_{n} = \infty, \;\; \forall\;\;
n \geq 1$ i.e. the average lifetime of the system is infinite. The
proof is simple: it is obvious that $\tau_{n} < \tau_{n+1}$,
therefore, it follows from (\ref{e4}) that $\sum_{i=1}^{n}
\delta_{i} < \tau_{1}$ for any $n$, and if $n \rightarrow \infty$,
than $\tau_{1}$ must be infinite. Since $\tau_{n} < \tau_{n+1}\;\;
\forall \; n \geq 1$, it is evident that $\tau_{n} = \infty, \;\;
\forall\;\; n \geq 1$, hence the statement is proven.

If $\lim_{n \rightarrow \infty} \sum_{i=1}^{n} \delta_{i} =
{\mathcal D} < \infty$, then one can find a finite real number $K$
such that $\prod_{i=1}^{n} (\lambda_{i}/\mu_{i}) < K$,
consequently
\[ \lim_{n \rightarrow \infty} (\tau_{n} -
\tau_{n+1})\;\prod_{i=1}^{n} \frac{\lambda_{i}}{\mu_{i}} = 0,\]
and it follows from this that
\begin{equation} \label{e5}
\tau_{1} = \sum_{i=1}^{\infty} \delta_{i}.
\end{equation}
Taking into account this relation  we can rewrite Eq. (\ref{e4})
into the form:
\[ (\tau_{n} - \tau_{n+1})\frac{1}{\rho_{n}} =
\sum_{i=n+1}^{\infty} \delta_{i},  \] where $\rho_{n}$ is defined
by (\ref{z41}). Introducing the notation
\begin{equation} \label{e6}
\chi_{n} = \sum_{i=n}^{\infty} \delta_{i},
\end{equation}
we have
\[ \tau_{n+1} = \tau_{n} + \rho_{n}\;\chi_{n+1}, \]
the solution of which is nothing else than
\[ \tau_{n} = \tau_{1} + \sum_{j=1}^{n-1} \rho_{j}\;\chi_{j+1}, \]
and by substituting $\tau_{1}$ and $\chi_{j+1}$ we obtain
immediately the equation (\ref{e0}).

\end{document}